\documentclass[aps,amsmath,amsfonts,11pt]{revtex4}
\usepackage{graphicx}
\usepackage{epstopdf}
\usepackage{epsfig,graphicx}
\usepackage[english]{babel}
\usepackage{amsfonts}
\usepackage{amsmath}
\usepackage{latexsym}
\usepackage{graphics,bm}
\usepackage{dcolumn}
\usepackage{bm}
\usepackage{rotating}

\begin{document}

\title{Fano profile in double cavity optomechanical system with harmonically bound mirrors}

\author{ Vaibhav N Prakash$^{1}$ and Aranya B Bhattacherjee$^{2}$ }

\address{$^{1}$School of Physical Sciences, Jawaharlal Nehru University, New Delhi-110067, India} 
\address{$^{2}$Department of Physics, Birla Institute of Technology and Science, Pilani,
Hyderabad Campus,  Hyderabad - 500078, India}

\begin{abstract}
 We investigate a double cavity optomechanical system(OMS) generating single and double Fano resonance(multi Fano). By altering a single parameter, the tunneling rate $g$ of the middle mirror, we are able to switch between single and double Fano line shapes. The first spectral line shape is stronger in the case of multi Fano than in the case of single Fano. Also the behaviour of steady state value of the displacement of the middle mirror with respect to $g$, heavily influences the behaviour of double Fano lines in our scheme. This tunability along-with using a single pump and signal/probe laser has an added advantage in situations where only low power consumption is available.   
\begin{description}
\item[Keywords]
Double Fano resonance, double-cavity OMS, optical sensors, optical switches.

\end{description}
\end{abstract}

\maketitle

\section{\label{sec:level1}Introduction\protect}

Fano resonance was first understood in the context of Rydberg atoms as an interference effect between discrete and continuum states \citep{Ugo}. Since then it has been observed in various physical processes involving bound states inside a continuum in atom-atom scattering. Besides atomic physics, Fano resonances can be found in nuclear physics, plasmonics and cavity optomechanics. A similar effect has been discovered in Mie scattering of small particles with negative dielectric susceptibility and weak dissipation rate \citep{Tribelsky}. Fano resonances are commonly observed when photons travel through different paths. In cavity optomechanics, they occur due to the constructive and destructive interference between two different pathways the photons travel to build the cavity field \citep{Agarwal}. The interaction of the discrete optomechanical ground state with broad continuum state in a $\Lambda$-type system results in an asymmetric line shape. This asymmetric structure of Fano line shape can then be conveniently used to produce fast all optical switches in communication networks \citep{Gao, Lassiter, Nozaki, Asadi} as opposed to the conventional switching components having long lorentzian tails. For similar reasons, Fano resonances can be used in optical sensors as it is very sensitive to changes in refractive index \citep{Feng, Zafar}. Fano minima has also been used for state transfer and transduction between microwave and optical photons \citep{Gu}. From a practical point of view it then becomes important to have greater tunability and efficiency to produce such line shapes so that they can be effortlessly incorporated into quantum devices.\\
On the other hand there has been a recent surge of interest in double cavity OMS. This is due to the greater tunability of membrane-in-the-middle devices over single cavity Fabry-Perot systems. Double cavity OMSs have been effectively used for ground state cooling \citep{Guo, Machnes}, entanglement \citep{Huan, Pinnard, Rarity} and transduction \citep{Fang} between microwave and optical photons. Hybrid optomechanical technologies(HOT) comprising of toroidal shaped whispering gallery mode (WGM) micro-resonators have been successfully used to induce non-reciprocity between communication channels \citep{Shang, Zhao}, an essential step towards quantum communication.\\
Here we report on the use of a double cavity optomechanical system(OMS) with two movable mirrors in producing single and double Fano resonances depending on the value of the tunneling rate g of the middle mirror/membrane. In our system we have used a single pump and probe to build an optical field inside a Fabry-Perot cavity. The double cavity OMS with two harmonically bound mirrors and a single pump and probe lasers seems to be energy efficient from a previously proposed system \citep{Agarwal}. Similar results have been achieved in interferometres having mirrors with different mechanical frequencies \citep{Farnaz} controlled by mechanical pumps \citep{Pramanik}. A major difference between these models and ours is the dependence on a single parameter, i.e the photon tunnelling rate $g$, to produce single and double Fano line-shapes. In this article we study the effect of changing $g$ on single and double Fano resonances. After giving a short description of the system Hamiltonian and establishing the Langevin equations in section II, we plot single and double Fano resonances in section III. In section IV we describe the behaviour of the two Fano line-shapes w.r.t $g$, followed by a discussion and conclusion in section V.

\section{\label{sec:level2}Basic Model}

We start with a scheme which we have previously discussed in \citep{Bhattacherjee}, shown in Fig.1. Cavity A is driven by an intense pump/control laser of frequency $\omega_{c}$ and has an average photon number $\overline{n}_a=<a^{\dagger}a>$ , where $a$ ($a^{\dagger}$) is the annihilation (creation) operator of optical mode confined in cavity A. 
\begin{figure}[h]
    \centering
    \includegraphics[scale=1.2]{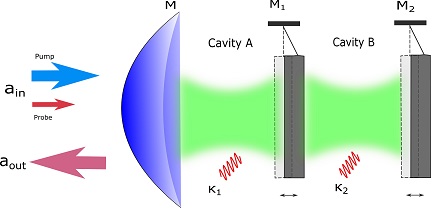}
    \caption{Schematic diagram of double cavity system with pump and probe from the left and output on the same side.Mirror $M$ is fixed while the mirrors $M_{1}$ and $M_{2}$ are movable. $\kappa_{1}$ and $\kappa_{2}$ are the mode decay rates of cavity A and cavity B respectively.}
    \end{figure}
The transparency of the middle mirror allows for the tunnelling of photons with rate $g$ into cavity B which has photons with average photon number $\overline{n}_b=<b^{\dagger}b>$  where $b$ ($b^{\dagger}$) is the annihilation (creation) operator of optical mode confined in cavity B. This leads to the coupling of the two mirrors $M_1$ and $M_2$  via radiation pressure forces from the cavity fields. In a frame rotating with frequency $\omega_c$, the system Hamiltonian can be written as,
\begin{equation}
\begin{aligned}
\hat{H}=\sum_{j=1}^{2}(\frac{\hat{p}_j^2}{2m_{j}}+\frac{1}{2}m_{j}\Omega_j^2\hat{x}_{j}^2)-\hbar\Delta_{1}\hat{a}^\dagger\hat{a}-\hbar\Delta_{2}\hat{b}^\dagger\hat{b}\\
-\hbar G_{1}\hat{x}_{1}\hat{a}^\dagger\hat{a}-\hbar G_{2}(\hat{x}_{2}-\hat{x}_{1})\hat{b}^\dagger\hat{b}\\
+\hbar g(\hat{a}^\dagger\hat{b}+\hat{a}\hat{b}^\dagger)+i\hbar\sqrt{\eta\kappa}\epsilon_{c}(\hat{a}^\dagger-\hat{a})\\+i\hbar\sqrt{\eta\kappa}\epsilon_p(\hat{a}^\dagger e^{-i\Omega t}-\hat{a} e^{i\Omega t}),
\end{aligned} 
\end{equation}
 where $m_1$ and $m_2$ are the 'bare' masses of mirrors $M_1$ and $M_2$ respectively. $\Omega_{j}(j$ = 1,2) is the mechanical frequency of the oscillator $M_{j}$. In our scheme both the movable mirrors vibrate with the same mechanical frequency i.e $\Omega_1=\Omega_2=\Omega_m$. We assume that the field in cavity B couples with equal strength to both mirrors $M_1$ and $M_2$. Also $\hat{x}_{1}$, $\hat{x}_{2}$ and
$\hat{p}_{1}$, $\hat{p}_{2}$ are the position and momentum operators following the commutation relations, $[\hat{x}_{j},\hat{p}_{j}]=i\hbar$ ($j=1,2$) for mirrors $M_{1}$ and $M_{2}$. $G_{1,2}$ denotes the cavity frequency shift per resonator displacement for $M_{1,2}$. It is related to single-photon optomechanical coupling strength by, $g_{0_{1,2}}=G_{1,2} x_{zp}$. Here $x_{zp}$ is the standard deviation of the zero point motion of the oscillator. Thus $G_{1}=\frac{\omega_{1}}{L_{1}}$ and $G_{2}=\frac{\omega_{2}}{L_{2}}$, where $\omega_{1}$($\omega_{2}$) are resonant frequencies of cavity A(B) and $L_{1,2}$ are the corresponding cavity lengths. It is to be noted that $G_{1,2}$ can be made very large as the effective length can be made very small \citep{Milburn}. Cavity A is driven by an external input laser field consisting of a strong control field and weak probe field denoted by $a_{in}(t)=\epsilon_{c}e^{-i\omega_{c}t}+\epsilon_p e^{-i\omega_p t}$ with field strengths $\epsilon_c$ and $\epsilon_p$ and frequencies $\omega_c$ and $\omega_p$ respectively. The field strengths are given as $\epsilon_c=\sqrt{P_{c}/\hbar \omega_{c}}$ and $\epsilon_p=\sqrt{P_{p}/\hbar \omega_{p}}$ where $P_{c}$ and $P_{p}$ are the control and probe field powers, respectively. Without loss of generality we assume that $\epsilon_c$ and $\epsilon_p$ are real. $\kappa_{ex}$ is the external decay rate between the i/o system and the optical cavity while $\kappa$ is the total decay rate. $\eta=\kappa_{ex}/\kappa$ is the coupling coefficient and is fixed in the critical coupling regime($\eta=0.5$) \citep{Zhang, Gallego}. Here $\Delta_j=\omega_{c}-\omega_{j}$(j=1,2), is the cavity detuning of cavity A ($j=1$) and B ($j=2$)  while $\Omega=\omega_p-\omega_c$ is the detuning of the probe field with respect to the control field frequency $\omega_c$. In our semi-classical description the operators are replaced by c-numbers while the noise terms are excluded. The Hamiltonian in Eq.1 gives rise to the following Heisenberg Langevin equation,

\begin{subequations}
\begin{equation} 
\dot{a}=i(\Delta_1+G_1x_{1})a-igb+\sqrt{\eta\kappa}\epsilon_c+\sqrt{\eta\kappa}\epsilon_pe^{-i\Omega t}-\frac{\kappa}{2}a,
\end{equation}
\begin{equation}
\dot{b}=i(\Delta_2+G_2[x_{2}-x_{1}])b-iga-\frac{\kappa}{2}b,
\end{equation}
\begin{equation}
\dot{x}_{j}=\frac{p_j}{m_j},
\end{equation}
\begin{equation}
\dot{p}_1=-m_1\Omega_1^2x_1-\hbar G_2b^\dagger b+\hbar G_1a^\dagger a-\frac{\gamma_1}{2}p_1,
\end{equation}
\begin{equation}
\dot{p}_2=-m_2\Omega_2^2x_2+\hbar G_2b^\dagger b-\frac{\gamma_2}{2} p_2,
\end{equation}
\end{subequations}

\section{\label{sec:level2}Fano resonances}
\begin{figure}[h]
    \centering
    \includegraphics[scale=0.85]{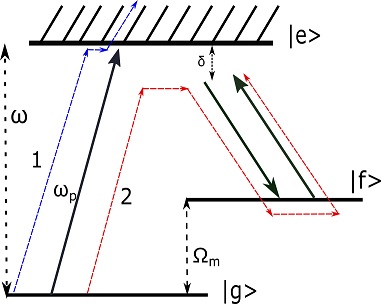}
    \caption{Schematic diagram of photon pathways}
    \end{figure}
Fano resonance can be observed in systems having a discrete state interacting with continuum state broadened by its decay rate. In cavity optomechanics it arises due to interference effects between one-photon absorption and Raman process making up a $\Lambda$-type system as illustrated in Fig.2. If $\omega$ is the cavity resonance frequency and $\omega_c$ is the frequency of the strong pump laser then the detuning parameter $\Delta=\omega_c-\omega$. The weak probe laser is detuned from $\omega_c$ such that it is near resonance with the cavity frequency. The mechanical oscillator with frequency $\Omega_m$ creates sidebands(Stokes and anti-Stokes) on the pump laser via Raman scattering. When the pump is red-detuned from the cavity frequency($\Delta\approx-\Omega_m$) such that the anti-Stokes sideband is resonant with the cavity frequency, photons from the probe laser can also be stimulated down to level $|f \rangle$ to produce phonons with frequency $\Omega_m$ and then back to the excited state $\ e \rangle$ with the help of the pump laser. Thus photons are excited inside the cavity via two different paths, labelled 1(one-photon/direct absorption process) and 2(Raman/indirect process). Just like the two slit experiment, since there is no way of knowing which slit the photons have come out, here there is no way of knowing which pathway the photons have traversed. This leads to interference effects. In fig.2 $\delta$ is a parameter which controls this interference. $\delta$ is the difference between the resonances of the probe field \citep{Agarwal}.  A complete destructive interference($\delta=0$) gives rise to OMIT(OptoMechanically Induced Transparency). The most general case arises due to a partial constructive and destructive interference($\delta\neq0$) between cavity photons and causes asymmetric Fano profiles.
\subsection{Single Fano} 
In cavity optomechanics, a $\Lambda$-type system shown in fig.2 can be formed using either single or multiple cavities separated by dielectric membranes. It has been shown \citep{Agarwal} that single Fano resonance can even occur in a single Fabry-Perot cavity provided that the anti-Stokes Raman field is offset from the cavity resonance frequency  such that their is a difference $\Omega_L$(or $\delta$ in our case) between the probe field and the anti-Stokes field. This gives rise to partial constructive and destructive interference causing asymmetric Fano line-shape. In a double cavity optomechanical system(OMS) too we can have single Fano provided the end mirrors are fixed. In that case the Hamiltonian in eq.1 will be modified;
\begin{eqnarray}
\begin{aligned}
\hat{H}=(\frac{\hat{p}_1^2}{2m_{1}}+\frac{1}{2}m_{1}\Omega_1^2\hat{x}_{1}^2)-\hbar\Delta_{1}\hat{a}^\dagger\hat{a}-\hbar\Delta_{2}\hat{b}^\dagger\hat{b}\\
-\hbar G_{1}\hat{x}_{1}\hat{a}^\dagger\hat{a}+G_2\hat{x}_{1}\hat{b}^\dagger\hat{b}\\
+\hbar g(\hat{a}^\dagger\hat{b}+\hat{a}\hat{b}^\dagger)+i\hbar\sqrt{\eta\kappa}\epsilon_{c}(\hat{a}^\dagger-\hat{a})\\+i\hbar\sqrt{\eta\kappa}\epsilon_p(\hat{a}^\dagger e^{-i\Omega t}-\hat{a} e^{i\Omega t}).
\end{aligned} 
\end{eqnarray}
Heisenberg Langevin eqs. 2(a-e) will also be modified according to eq.3; 
\begin{subequations}
\begin{equation} 
\dot{a}=i(\Delta_1+G_1x_1)a-igb+\sqrt{\eta\kappa}\epsilon_c+\sqrt{\eta\kappa}\epsilon_pe^{-i\Omega t}-\frac{\kappa}{2}a,
\end{equation}
\begin{equation}
\dot{b}=i(\Delta_2-G_2x_1)b-iga-\frac{\kappa}{2}b,
\end{equation}
\begin{equation}
\dot{x}_1=\frac{p_1}{m_1},
\end{equation}
\begin{equation}
\dot{p}_1=-m_1\Omega_1^2x_1-\hbar G_2b^\dagger b+\hbar G_1a^\dagger a-\frac{\gamma_1}{2}p_1,
\end{equation}
\end{subequations}
where $x_1$ is the displacement of the middle mirror/membrane. The above equations are almost similar to the ones in \citep{Zhang}, where a toroidial WGM microresonator was studied instead of a Fabry-Perot cavity. Using standard mean field and linearization approach \citep{Aspelmeyer} and breaking the above c-numbers(eqs. 4(a-d)) into their respective Fourier components (Y=$\sum_{n=1,2}S_ne^{\pm i\Omega t}$)(Appendix A), we arrive at the anti-stokes field in cavity A.
\begin{equation}
    \begin{aligned}
    A_1^-=-\frac{(gG_2|\overline{b}|q_1+i D_3G_1|\overline{a}|q_1+D_3\sqrt{\eta\kappa}\epsilon_p)}{D_1D_3+g^2},
    \end{aligned}
\end{equation}
where $D_1=\Theta_1+i \Omega$, $D_2=\Theta_1-i \Omega$, $D_3=\Theta_2+i \Omega$, $D_4=\Theta_2-i \Omega$ and $\Theta_j=i\overline{\Delta}_j-\kappa/2, (j=1,2)$.
Since we are using a single input-output port, we can measure only the backward reflection coefficient of the input probe field. The standard input-output relation leads to;
\begin{equation}
\begin{aligned}
a_{out}=C_{cb}e^{-i\omega_c t}+C_{pb}e^{-i\omega_p t}-\sqrt{\eta\kappa}A_1^+e^{-i(2\omega_c-\omega_p)t},
\end{aligned}
\end{equation}
where $C_{cb}=\epsilon_c-\sqrt{\eta\kappa} \overline{a}$, $C_{pb}=\epsilon_p-\sqrt{\eta\kappa}A_1^-$, are the complex coefficients for steady state and backward reflection, respectively. 
Using $C_{pb}$ we calculate the normalized backward reflection coefficient, $T_b=|C_{pb}/\epsilon_p|^2$(Appendix A, eq.A2). Since the pump laser is red-detuned by the mechanical frequency $\Omega_m$, from eq.6, the Stokes field, with coefficient $A_1^+$, will be off-resonant by $2\omega_c-\omega_p$ from the cavity. Hence we do not show it here but it is straightforward to calculate.
\begin{figure}[h]
    \centering
\begin{tabular}{cc}
\includegraphics [scale=0.55]{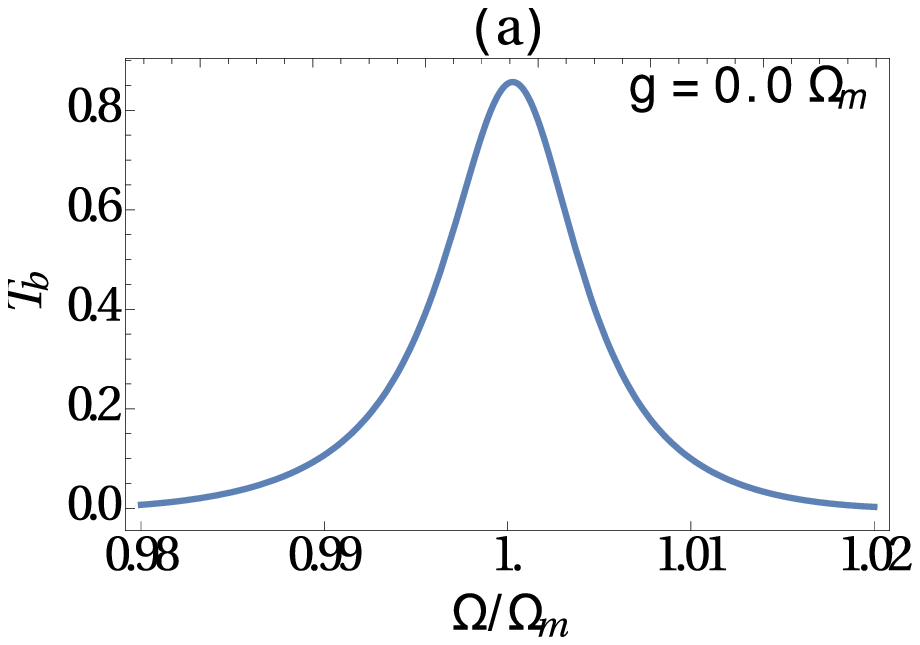} \includegraphics [scale=0.55] {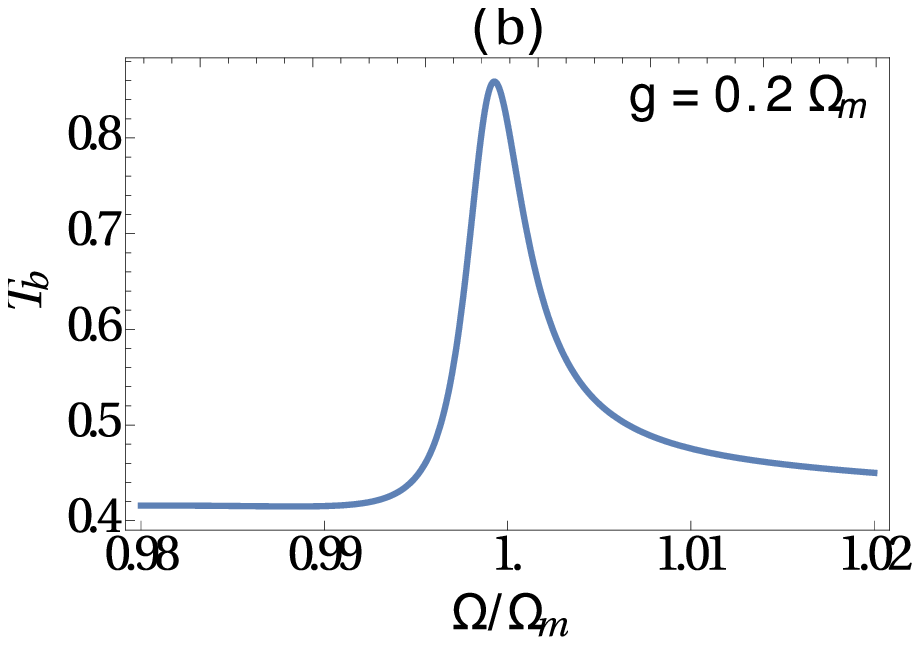}\\
\includegraphics [scale=0.55]{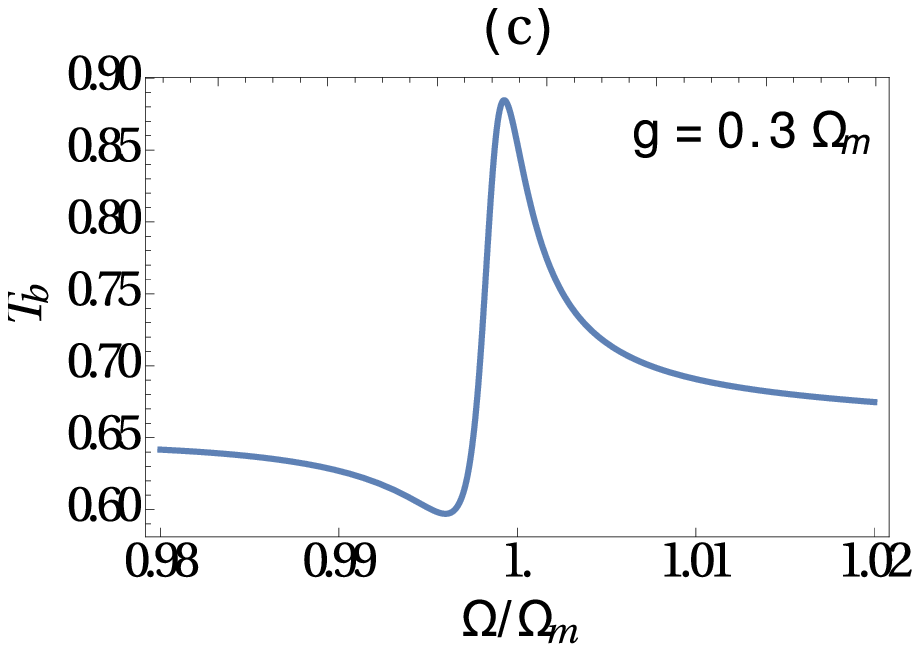} \includegraphics [scale=0.55] {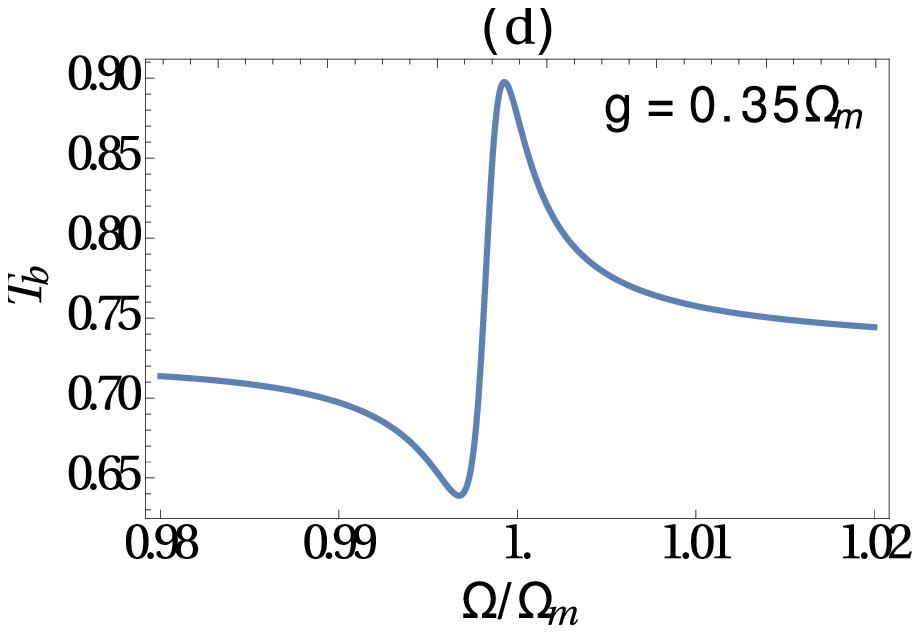}\\
\includegraphics [scale=0.55]{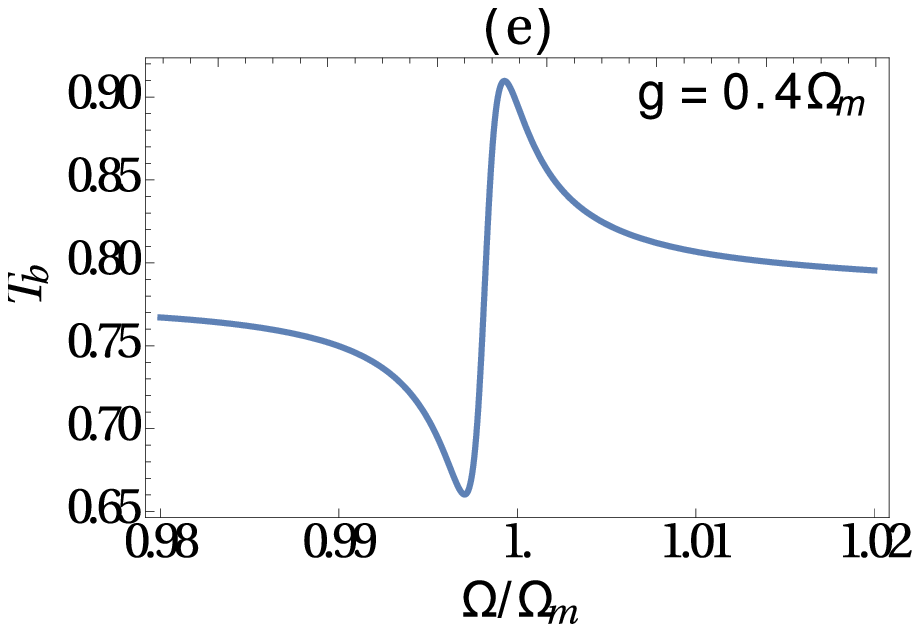} \includegraphics [scale=0.55] {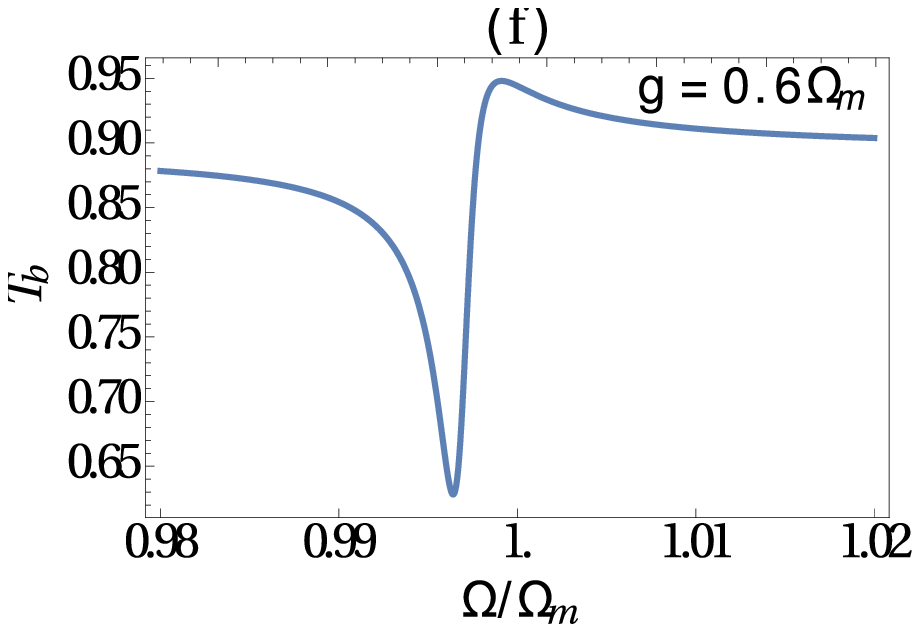}\\
 \end{tabular}
\caption{(Color online) Backward reflection coefficient for membrane-in-middle resonator with fixed end mirrors.  $m_1=20ng$, $P_c=1mW$, $G_{1}=2\pi\times 13GHz/nm$, $G_{2}=2\pi\times 13GHz/nm$, $\gamma_{1}=\gamma_{2}=2\pi\times41kHz$, $\kappa=2\pi\times15MHz$, $\Delta_{2}=\Delta_1=-\Omega_{m}$ and $\Omega_1=\Omega_m=2\pi\times51.8MHz$.}
\end{figure}
Fig.3 gives the normalized backward reflection coefficient $T_b$ for different values of tunneling rate g and $\Omega/\Omega_m \in$ [0.98, 1.02]. This is done to emphasize changes in the complete destructive interference(leading to OMIT) of the cavity field to a constructive and destructive interference(leading to asymmetric Fano line shapes). OMIT occurs at $g/\Omega_m=0$ (fig.3a) since both the probe and the anti-Stokes Raman field completely destructively interfere at cavity resonance. For $0.0<g/\Omega_m\leq0.4$(fig. 3(b-e) $T_b$ intensity changes gradually with peak around 0.9. For $g/\Omega_m\sim0.6$ (fig. 3f) the peak drastically increases to around 1.0.
\newpage
\subsection{Multi Fano}
\begin{figure}[h]
\centering
\begin{tabular}{cc}
\includegraphics [scale=0.55]{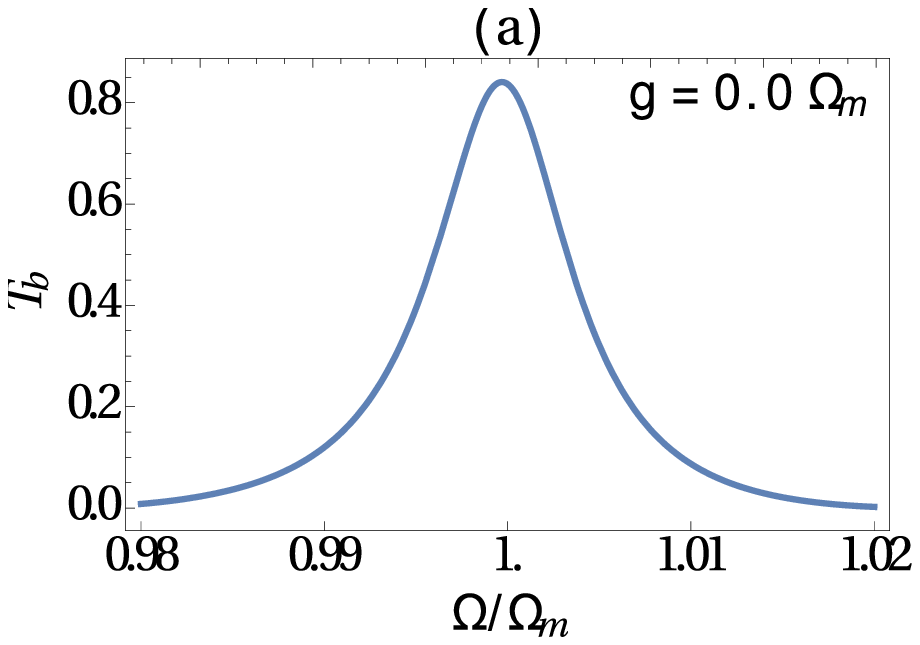} \includegraphics [scale=0.55] {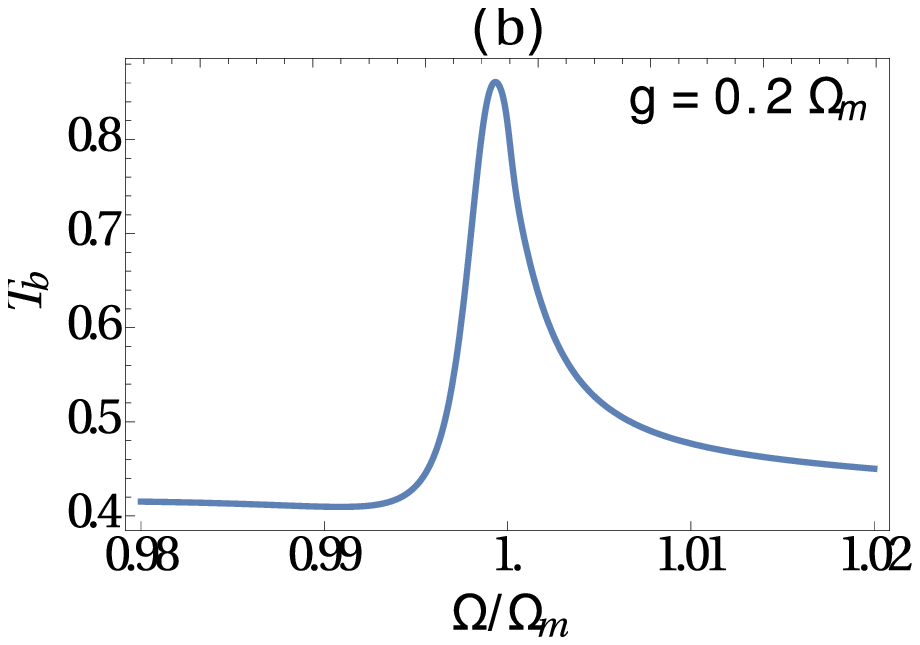}\\
\includegraphics [scale=0.55]{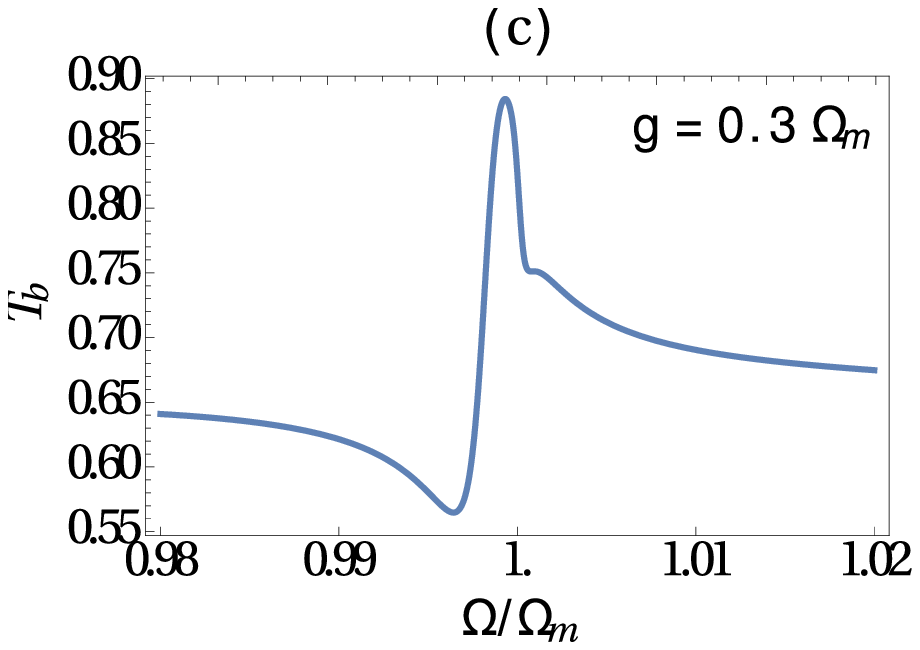} \includegraphics [scale=0.55] {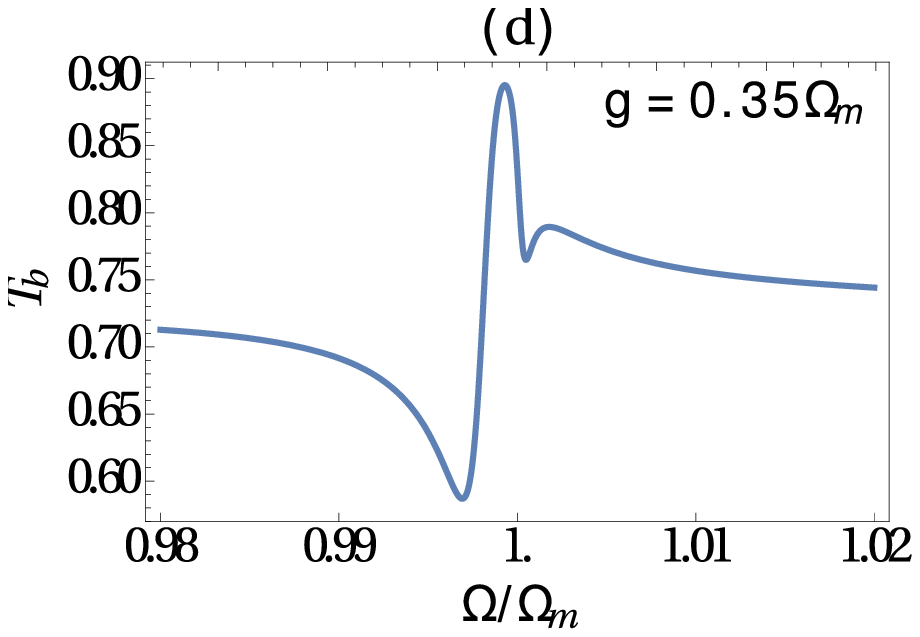}\\
\includegraphics [scale=0.55]{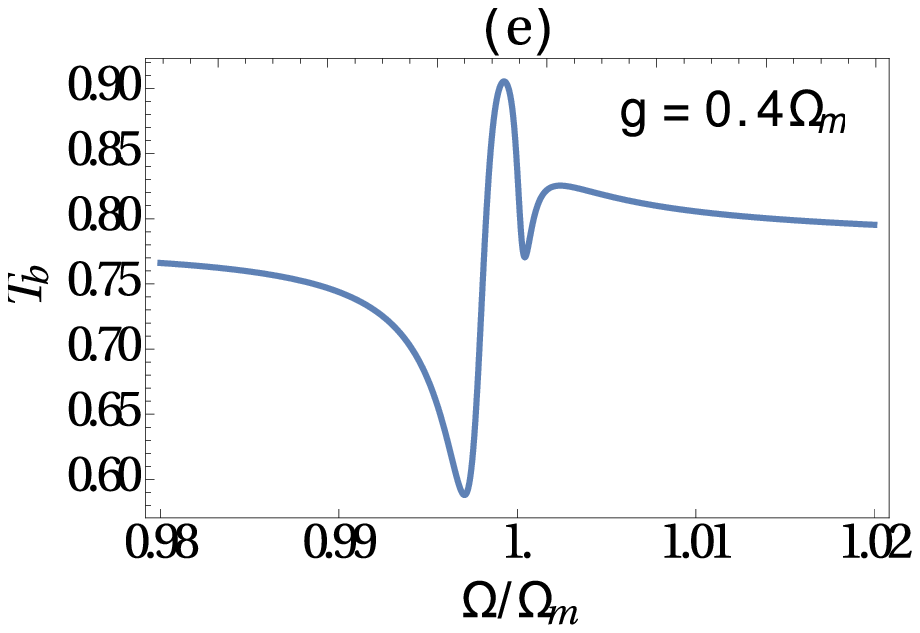} \includegraphics [scale=0.55] {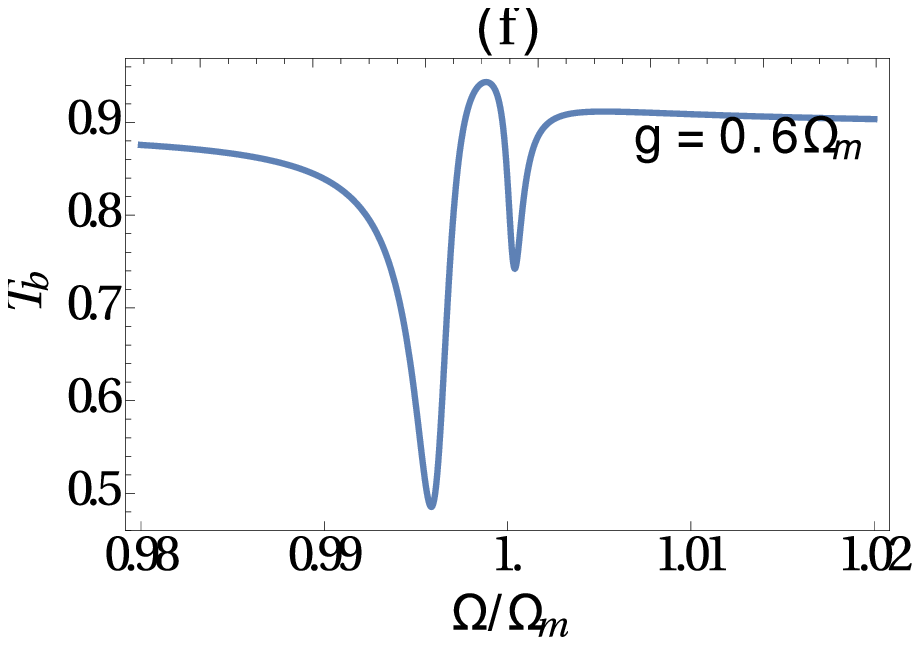}\\
 \end{tabular}
 \caption{(Color online) Backward reflection coefficient for membrane-in-middle resonator with end mirrors fixed.  $m=20ng$, $P_c=1mW$, $G_{1}=2\pi\times 13GHz/nm$, $G_{2}=2\pi\times 13GHz/nm$, $\gamma_{1}=\gamma_{2}=2\pi\times41kHz$, $\kappa=2\pi\times15MHz$, $\Delta_{2}=\Delta_1=\Omega_{m}$ and $\Omega_1=\Omega_2=\Omega_m=2\pi\times51.8MHz$.  }
\label{fig:3}
\end{figure}
Fano profiles depend on the number of dressed states and resonances of the probe laser. For example a single dressed state with single resonance gives rise to a single OMIT. On the other hand the occurrence of a double OMIT has been recently shown \citep{Shi-Chao} in a piezo-mechanical system coupled with an OMS. The piezo-mechanical coupling, forms two dressed states with two resonances, changing a single OMIT into a double OMIT. A shift in the detuning of the coupling laser by an amount $\delta$ gives rise to more number of resonances than dressed states leading to asymmetric Fano profiles. Here we specifically show the existence of double asymmetric Fano-line shapes in a double cavity OMS. The system dynamics are defined by eqn.2(a-e) and following the same procedure given in section III-A, the anti-stokes field component $A_1^-$ is given as;
\begin{equation}
\begin{aligned}
A_1^-&=\frac{gG_2(q_{2}-q_{1})\overline{b}-iD_{3}G_1q_{1}\overline{a}-D_{3}\sqrt{\eta\kappa}}{D_1D_3+g^2}.
\end{aligned}
\end{equation}
The normalized backward reflection coefficient $T_b=|C_{pb}/\epsilon_p|^2$ as defined below eq.6, is given in Appendix A(eq. A4). We plot $T_b$ in fig.4(a-f). The x-axis range here is again $\Omega/\Omega_m \in$ [0.98, 1.02] so as to emphasize changes in Fano profiles. As in fig.3a, in fig.4a we have a single OMIT structure at $g/\Omega_m=0.0$. This system does not show double OMIT since at $g/\Omega_m=0.0$ the system turns into a single cavity OMS with complete destructive interference of the Raman anti-stokes and the probe field at cavity resonance with response function ($\epsilon_T=\sqrt{\eta\kappa} A_1^-/\epsilon_p$) of the OMS being,
\begin{equation}
    \begin{aligned}
    \epsilon_T=\frac{\eta\kappa}{\frac{\kappa}{2}-iy+\frac{\beta}{\frac{\gamma_m}{2}-iy}}.
    \end{aligned}
\end{equation}
We note that instead of having a large parameter space, we have fixed all the parameters except for the tunneling rate $g$. This makes the system experimentally easy to access. Comparing fig.3 and 4 we see that the Fano line-shapes are roughly similar upto around $g/\Omega_m=0.2$. At around  $g/\Omega_m=0.3$ we observe the first signs of single Fano going into double Fano. At $g/\Omega_m=0.4$ we clearly see the separation of the two distinct asymmetric Fano line-shapes. The first Fano line-shape for double Fano case is more sharp than that for the single Fano case(fig.3). This means easier switching when implemented in all-optical switches. We also note that the difference between strength of dips(y-axis) for the two Fano lines remain more or less constant with increasing $g$. Hence a sharper first Fano line also means a sharper second Fano line.\\
\section{Fano profile behaviour}
Next we try to model the behaviour of the double Fano profile. Using analytical tools(Mathematica), we measure the distance($\Omega/\Omega_m$) between the two Fano line-shapes in the backward reflection($T_b$) plot. The dependency of the measured separation values($\Omega/\Omega_m$) between the two line-shapes is plotted with respect to $g$ for 0.4$\leq g/\Omega_m\leq$1. We choose this interval due to prominence of the weaker line-shape. For comparison, in fig.5a, we also plot scaled up steady state values $\overline{x}_1$ and $\overline{x}_2$ for displacements of mirrors $M_1$ and $M_2$ respectively.
\begin{subequations}
\begin{equation}
\overline{x}_1=\frac{\hbar(G_1\mid\overline{a}\mid ^2-G_2\mid\overline{b}\mid ^2)}{m_1\Omega_1^2},
\end{equation}
\begin{equation}
\overline{x}_2=\frac{\hbar G_2\mid\overline{b}\mid^2}{m_2\Omega_2^2}.
\end{equation}
\end{subequations}
\begin{figure}[h]
\begin{tabular}{c}
\includegraphics [scale=1.1]{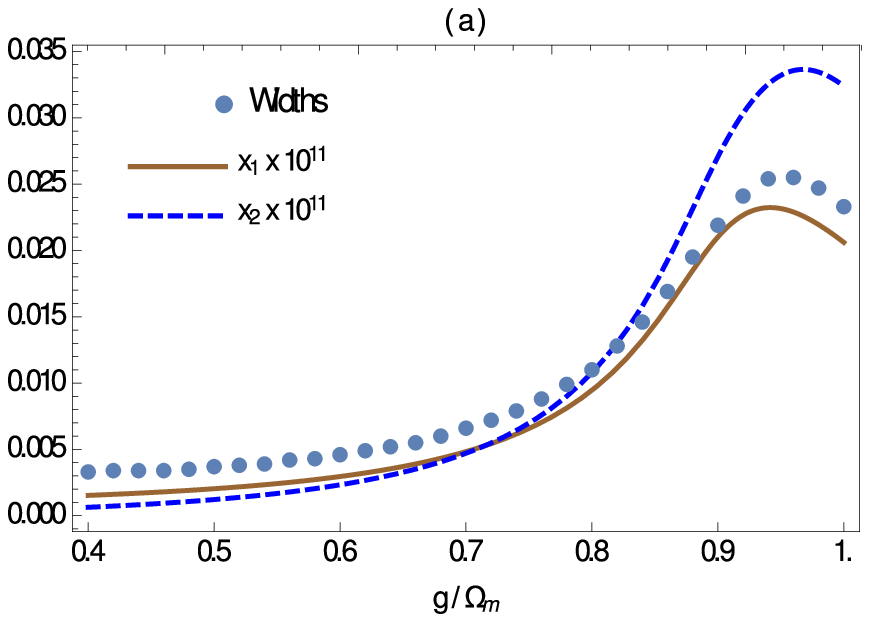}\\
 \includegraphics [scale=1.0] {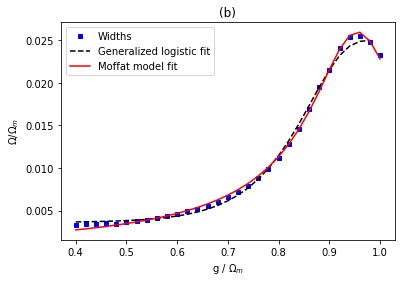}\\
\end{tabular}
 \caption{(Color online) (a) Calculated steady state values $\overline{x}_1$ and $\overline{x}_2$ scaled up by an order of $10^{11}$ compared with numerically calulated values of the widths($\Omega/\Omega_m$) between Fano line-shapes. (b) Fits of two different functions with the measured values.}
\label{fig:5}
\end{figure}
It can be seen in fig.5a that the behaviour of measured values closely mimics that of $\overline{x}_1$. It should be noted that large separation value amounts to well-resolved Fano lines. From fig.5a it can be seen that the transmitivity(or tunnelling rate $g$) of the middle mirror $M_1$ should be relatively high to guarantee well-resolved Fano lines. There is a rapid increase for $0.6\leq g/\Omega_m\leq 0.9$ saturating at $g/\Omega_m=0.95$ before decreasing till $g/\Omega_m=1$. We use two functions to fit the data as shown in fig.5b, a generalized logistic function and the Moffat function \citep{moffat},
\begin{subequations}
\begin{equation}
    Y_{gen}(x) = a + \frac{c}{1+T e^{{-B(x-M)}^{\frac{1}{T}}}},
\end{equation}
\begin{equation}
    Y_{moff}(x) = A (1+(\frac{x-\mu}{\sigma})^2)^{-\beta},
\end{equation}
\end{subequations}
with appropriate fit parameters. A generalised logistic function is a cumulative function depicting, among many things, diffusion. On the other hand Moffat function is a point spread function(PSF) used in astrophysics to fit 'seeing-limited' images of point sources \citep{trujillo}, a blurring of light caused by the earth's atmosphere. Hence it also becomes a measure of diffusion of light emanating from a point source registering on a photographic plate after passing through the earth's atmosphere. Both the fits give low $\chi^2/d.o.f$(degree of freedom) values of the order of $10^{-4}$ but the Moffat function gives a better fit as seen from fig.5b. Even though both functions have very different functional forms, from the fits we argue that the separation between the two Fano profiles can be controlled by controlling the diffusion of photons from cavity A to cavity B. Hence in our scheme the tunnelling rate $g$ of the middle mirror $M_1$ becomes an important parameter.
\section{\label{sec:level2}Discussion and Conclusion}
\begin{figure}[h]
\centering
\includegraphics [scale=0.72]{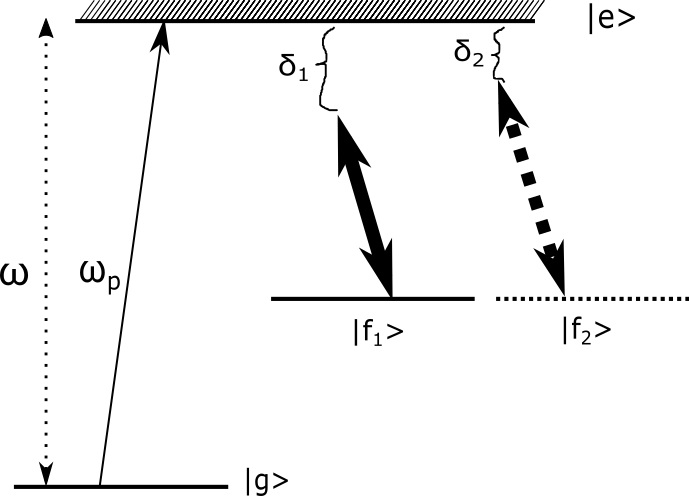}
 \caption{(Color online) Schematic diagram for single and double Fano resonance.}
\label{fig:6}
\end{figure}
Here we do not delve into quantitatively understanding the results given in Fig.3 and Fig.4 following equations for the normalized backward reflection coefficients, but try to intuitively understand them. The pump laser is red-detuned from cavity A by an amount $\Omega_m$. 
The middle resonator allows photons into the second cavity with tunnelling rate $g$. Due to radiation pressure on mirror $M_2$ and since mirror $M_2$ has the same natural frequency $\Omega_m$ as $M_1$, we get another state $|f_{2} \rangle$ $\textit{\textbf{degenerate}}$ with $|f_{1} \rangle$ as shown in Fig.6. The simultaneous coupling of photons in cavity A and B with mirrors $M_1$ and $M_2$ gives rise to shifts $\delta_1$ and $\delta_2$, respectively. In our scheme a single OMIT occurs when $\delta_1$=$\delta_2$=0 i.e when only cavity A and mirror $M_1$ is activated($g/\Omega_m=0$). Fig.7 gives the strength of fields in both the cavities. At lower values of the tunneling rate $g$, the field strength in cavity B is much smaller than that in cavity A. This means that it cannot substantially affect the motion of mirror $M_2$ via radiation pressure. 
\begin{figure}[h]
\centering
\includegraphics [scale=0.9]{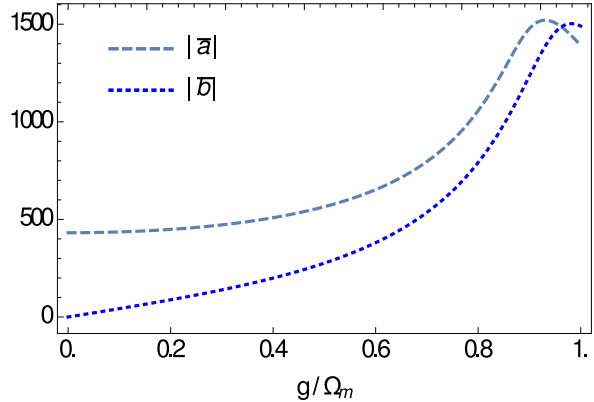}
 \caption{(Color online) Photon intensity in cavity A and B.}
\label{fig:7}
\end{figure}
Thus state $|f_{2} \rangle$ is practically non-existent and $\delta_1\neq$0 but $\delta_2$=0, implying only single Fano. Hence Fano profiles in Fig. 3 and 4 look pretty similar upto around $g/\Omega_m=0.2$. Once the value of $g/\Omega_m\geq0.3$, the optical field in cavity B becomes strong enough to influence motion of $M_2$ and hence $\delta_2\neq0$. This might be the reason we start observing a shift from single to double Fano around $g/\Omega_m=0.3$ in fig.4. The line-shapes become prominent for $g/\Omega_m\geq0.4$ since the field strengths in cavity A and B increases with $g$.\\
The saturation of different values in fig.5a before $g/\Omega_m=1$ is due to the approximations(for e.g., truncation of higher order optomechanical coupling, mean field approach and rotating wave approximation) used in determining these values. Although these are good approximations for large cavity field strengths(fig.7), they do have their short comings. For instance, in single photon optomechanics nonlinear photon-phonon interaction becomes important and one can no longer assume linear response of oscillating mirrors to solve the equations of motion \citep{Nunnenkamp}. In the single photon strong coupling regime one also has to include higher order expansion of cavity frequency $\omega(x)$, leading to higher order optomechanical coupling rate.\\
In conclusion, we have successfully shown the presence of double Fano profiles in our scheme. We have also shown that the same scheme can be used to generate single and double Fano lines-shapes by simply controlling the tunnelling rate $g$ of the middle mirror $M_1$. The behaviour of the separation between the line-shapes closely mimics the behaviour of the steady state displacement of $M_1$ with photons diffusing from cavity A to cavity B. The first Fano line has a larger dip in the multi Fano case than in the single Fano for the same value of $g/\Omega_m$. The double Fano line-shapes also become resolved for 0.6$\leq g/\Omega_m\leq$0.95. Hence our scheme can be implemented for sensitive devices in need of sharp spectral lines like in all-optical switching and quantum sensors.

\appendix
\section{}
The Fourier components for cavity and mechanical c-numbers given in eq.2 and 4 are,
\begin{subequations}
\begin{equation}
\begin{aligned}
\delta{a}(t)=A_1^-e^{-i\Omega t}+A_1^+e^{i\Omega t},
\end{aligned}
\end{equation}
\begin{equation}
\begin{aligned}
\delta{b}(t)=B_1^-e^{-i\Omega t}+B_1^+e^{i\Omega t},
\end{aligned}
\end{equation}
\begin{equation}
\begin{aligned}
\delta{x_1}(t)=q_1 e^{-i\Omega t}+q_1^*e^{i\Omega t}\hspace{0.2cm},
\end{aligned}
\end{equation}
\begin{equation}
\begin{aligned}
\delta{x_2}(t)=q_2 e^{-i\Omega t}+q_2^*e^{i\Omega t}\hspace{0.2cm}.
\end{aligned}
\end{equation}
\end{subequations}
Single Fano(assuming $D_1\neq D_3$, $D_2\neq D_4$ and $G_1=G_2=G$ ) :\\
\begin{equation}
\begin{aligned}
    T_b=\left|1-\eta\kappa\left[\frac{ig^2G|\overline{b}|^2-iGD_3^2|\overline{a}|^2-gD_3G(\overline{a}^*\overline{b}+\overline{a}\overline{b}^*)}{(D_1D_3+g^2)^2(C_1'+C_2'+C_3')}-\frac{D_3}{(D_1D_3+g^2)}\right]\right|^2,
\end{aligned}
\end{equation}\\
where,
\begin{subequations}
\begin{equation}
    C_1'= \frac{-g G (\overline{a}^*\overline{b}+\overline{a}\overline{b}^*)+i G(D_3|\overline{a}|^2+D_1|\overline{b}|^2)}{(D_1 D_3 + g^2)},
\end{equation}
\begin{equation}
    C_2'= \frac{-g G(\overline{a}^*\overline{b}+\overline{a}\overline{b}^*)-i G(D_4^*|\overline{a}|^2+D_2^*|\overline{b}|^2)}{(D_2^* D_4^* + g^2)},
\end{equation}
\begin{equation}
    C_3'=-\frac{1}{\hbar G\chi_1(\Omega)}.
\end{equation}
\end{subequations}
$\chi_1(\Omega)$ is the mechanical susceptibility of mirror $M_1$.\\\\
Double Fano(assuming $D_1=D_3$, $D_2=D_4$ and $G_1\neq G_2$) :\\
\begin{equation}
\begin{split}
 T_{b}=&|1-\eta\kappa[\frac{\frac{i g^2 G_{2}^2 \left(1-\frac{A}{B}\right)
   (1+C_{11}+C_{22})}{G_{1}}|\overline{b}|^2-i G_{1} D_{1}^2|\overline{a}|^2 -D_{1} g G_{2} ( (1-\frac{A}{B})\overline{a}^*\overline{b}+
   (1+C_{11}+C_{22})\overline{a}\overline{b}^*)}{(D_{1}^2+g^2)^2
   (C_{1}+C_{2}+C_{3})}\\\\&+\frac{ig^2 G_{2} |\overline{b}|^2 }{B
   (D_{1}^2+g^2)^2}-\frac{D_{1}}{(D_{1}^2+g^2)}]|^2,
 \end{split}
\end{equation}
where,
\begin{subequations}
\begin{equation}
 \begin{aligned}
 A=-\frac{g G_{1}\overline{a}\overline{b}^* +i D_{1} G_{2}|\overline{b}|^2 }{D_{1}^2+g^2}-\frac{g G_{1}\overline{a}^*\overline{b} -i  G_{2} D_{2}^*|\overline{b}|^2
  }{\left(D_{2}^*\right)^2+g^2},
\end{aligned}
\end{equation}
\begin{equation}
 \begin{aligned}
 B=\frac{iG_{2} D_{2}^* |\overline{b}|^2 }{\left(D_{2}^*\right)^2+g^2}-\frac{i D_{1}
   G_{2}|\overline{b}|^2 }{D_{1}^2+g^2}-\frac{1}{\hbar G_{2} {\chi_{2}(\Omega)} },
\end{aligned}
\end{equation}
\begin{equation}
 \begin{aligned}
 C_{1}=\frac{-g G_{2} \left(\left(1-\frac{A}{B}\right)\overline{a}\overline{b}^*+\overline{a}^*\overline{b}\right)+\frac{i G_{2}^2
   \left(1-\frac{A}{B}\right)|\overline{b}|^2 D_{2}^*}{G_{1}}+iG_{1}|\overline{a}|^2 D_{2}^*}{\left(D_{2}^*\right)^2+g^2},
\end{aligned}
\end{equation}
\begin{equation}
 \begin{aligned}
 C_{2}=\frac{-g G_{2} \left(\left(1-\frac{A}{B}\right)\overline{a}^*\overline{b} +\overline{a}\overline{b}^*\right)-\frac{iG_{2}^2\left(1-\frac{A}{B}\right)|\overline{b}|^2 D_{1} 
   }{G_{1}}-i G_{1}|\overline{a}|^2 D_{1} }{D_{1}^2+g^2},
\end{aligned}
\end{equation}
\begin{equation}
 \begin{aligned}
 C_{3}=-\frac{1}{ \hbar G_{1} \chi_{1}(\Omega)},
 \end{aligned}
\end{equation}
\begin{equation}
 \begin{aligned}
 C_{11}=\frac{g G_{1}\overline{a}^*\overline{b} +iG_{2} D_{1} |\overline{b}|^2}{B \left(D_{1}^2+g^2\right)},
 \end{aligned}
\end{equation}
\begin{equation}
 \begin{aligned}
 C_{22}=\frac{g G_{1}\overline{a}\overline{b}^* -i  G_{2} D_{2}^*|\overline{b}|^2}{B \left(\left(D_{2}^*\right)^2+g^2\right)}.
 \end{aligned}
\end{equation}
\end{subequations}\\
$\chi_2(\Omega)$ is the mechanical susceptibility of mirror $M_2$.\\
 
\end{document}